\documentclass[useAMS,usenatbib]{mn2e}
\usepackage{graphicx}

\newcommand{\etal}{et al.}

\title[The Core-like Nature of HST-1]{The Core-like Nature of HST-1 in the M87 Jet}
\author[Y. J. Chen, G.-Y. Zhao and Z.-Q. Shen]{
Y. J. Chen$^{1,2}$\thanks{E-mail: cyj@shao.ac.cn}, G.-Y. Zhao$^{1,3}$ and  Z.-Q. Shen$^{1,2}$\\
$^{1}$Key Laboratory for Research in Galaxies and Cosmology, Shanghai Astronomical Observatory, Chinese Academy of
Sciences,\\80 Nandan Road Shanghai 200030, China\\
$^{2}$Key Laboratory of Radio Astronomy, Chinese Academy of Sciences\\
$^{3}$Graduate School of the Chinese Academy of Sciences, Beijing 100039, China}

\begin{document}

\date{}

\pagerange{\pageref{firstpage}--\pageref{lastpage}} \pubyear{}

\maketitle

\label{firstpage}

\begin{abstract}
We investigate the total flux density, spectral, polarization, and Faraday rotation variability of HST-1 in the M87 jet during the outburst from 2003 to 2007 through multi-epoch VLA observations at 8, 15, and 22 GHz. Contrary to the general case for blazars, the flux densities of HST-1 rise earlier at lower frequencies from radio to X-ray, and the spectra are softening with the growth of outburst, indicating that the newly emerging subcomponents within HST-1 have relatively steep spectra. In particular, the intrinsic EVPA varies monotonically by $\sim90^\circ$ at the 3 wavebands during the period, and all but the stationary subcomponent in the eastern end of HST-1 move downstream superluminally deviating divergently from the overall jet direction, with the motion of the outmost subcomponent bending from one side of the jet axis to another. These strongly argue for the presence of helical magnetic fields around HST-1, which is also supported by the fact that the subcomponents might be accelerated in this region. The fractional polarization is relatively low in the rising stage, and in the decaying stage the polarization levels are almost comparable at the 3 wavebands. In view of the quite large RM values, Faraday rotation is expected to occur dominantly external to HST-1 in the decaying stage, which is well supported by the presence of diffuse emission around HST-1, and consistent with the scenario that RM decrease gets slower with time.
\end{abstract}


\section[]{Introduction}

Great progress has been made in understanding jet acceleration and collimation. Evidence is found that the optical electric vector positional angle (EVPA) of a newly formed feature rotates while it moves through a region upstream of the radio cores in BL Lac and PKS 1510-089, which is attributed to helical magnetic fields accelerating and focusing the jet in that zone \citep{mar08,mar10}. In these two cases, the accelerating and focusing process lasts until the feature encounters a region of turbulence, where the magnetic fields become chaotic. Whereas further attempts to investigate the process are still required, the jets in quite a few sources are found to be accelerated downstream of the core (e.g., \citealt{jor05}), and helical magnetic fields are argued to be present within the jets in many sources by identifying the Faraday rotation measure (RM) gradients across the jet downstream from the core (e.g., \citealt{zav05,gab08}). It is therefore likely that large-scale helical magnetic fields exist in jets, and play an important role in jet acceleration and collimation downstream of the core as well, as suggested by \cite{lyu05}.

M87 is one of the closest radio galaxies with a distance of 16.7~Mpc, which is active from radio to $\gamma$-ray waveband (e.g., \citealt{acc09}). HST-1 in M87 jet was defined to be a feature or component separated by $\sim0.86''$ from the central black hole, which is resolved into several `subcomponents' at mas resolution with the most east one stationary and some others moving down the jet superluminally \citep{che07}. It seems that new subcomponents are ejected frequently from the stationary one. The flux density monitoring at multi-wavebands shows that a major outburst of HST-1 occurred from 2003 to 2007, of which the radiation was regarded to be synchrotron from radio to X-ray \citep{har03,har09}. Along with a $\gamma$-ray flare occurred in the period \citep{aha06}, HST-1 was argued to show  significant blazar-like properties (e.g., \citealt{har08}), of which the polarization nature has yet been rarely investigated.

Here we present the multi-epoch full polarization results of HST-1
in M87, which are derived from VLA observations at 8, 15, and 22 GHz
in the same \emph{A} array. \S2 describes observations and data reduction, followed by the results and discussion in \S3. Finally, a summary is given in \S4.

\section[]{Observations and data reduction}
We collected the multi-frequency full polarization data from NRAO Data Archive \footnote{The National Radio Astronomy Observatory is a facility of the National Science Foundation operated under cooperative agreement by Associated Universities, Inc.}, which were observed with the VLA A array including 8 epochs spanning the major outburst period from 2003 to 2007 to roughly keep consistency in resolution and sensitivity from epoch to epoch. The observations utilized two adjacent 50 MHz wide intermediate frequencies at all the three wavebands.

Data calibration were performed in AIPS \citep{bri94} with flux density and EVPA calibrated using 3C 286. Phases were initially calibrated through a nearby point source 1224+035. Instrumental polarization and EVPA corrections were obtained by using `PCAL' and `RLDIF', respectively. After the initial amplitude and phase calibration, the data were then read into DIFMAP \citep{she94} for
imaging, with the resultant full polarization images loaded back into AIPS for model fitting using the task `JMFIT'. Since HST-1 is not well separated from the core at 8 GHz , the whole core region containing HST-1 is taken to be fitted with 3 elliptical components at 8 GHz, and 4 elliptical components at 15 and 22 GHz, and at all epochs the feature HST-1 is unambiguously identified. Note that since HST-1 is almost point-like at the VLA resolution, the effect of different beam size at the 3 frequencies on the results could be neglected. When model-fitting to the Stokes `Q' and `U', we fixed the size and position of each component derived from fits to the total intensity image to ensure that the resultant polarized intensity came from the same emission region (see Fig.~\ref{fig1}).

\section[]{Results and discussion}

\begin{figure}
\centerline{\includegraphics[width=6cm, angle=0]{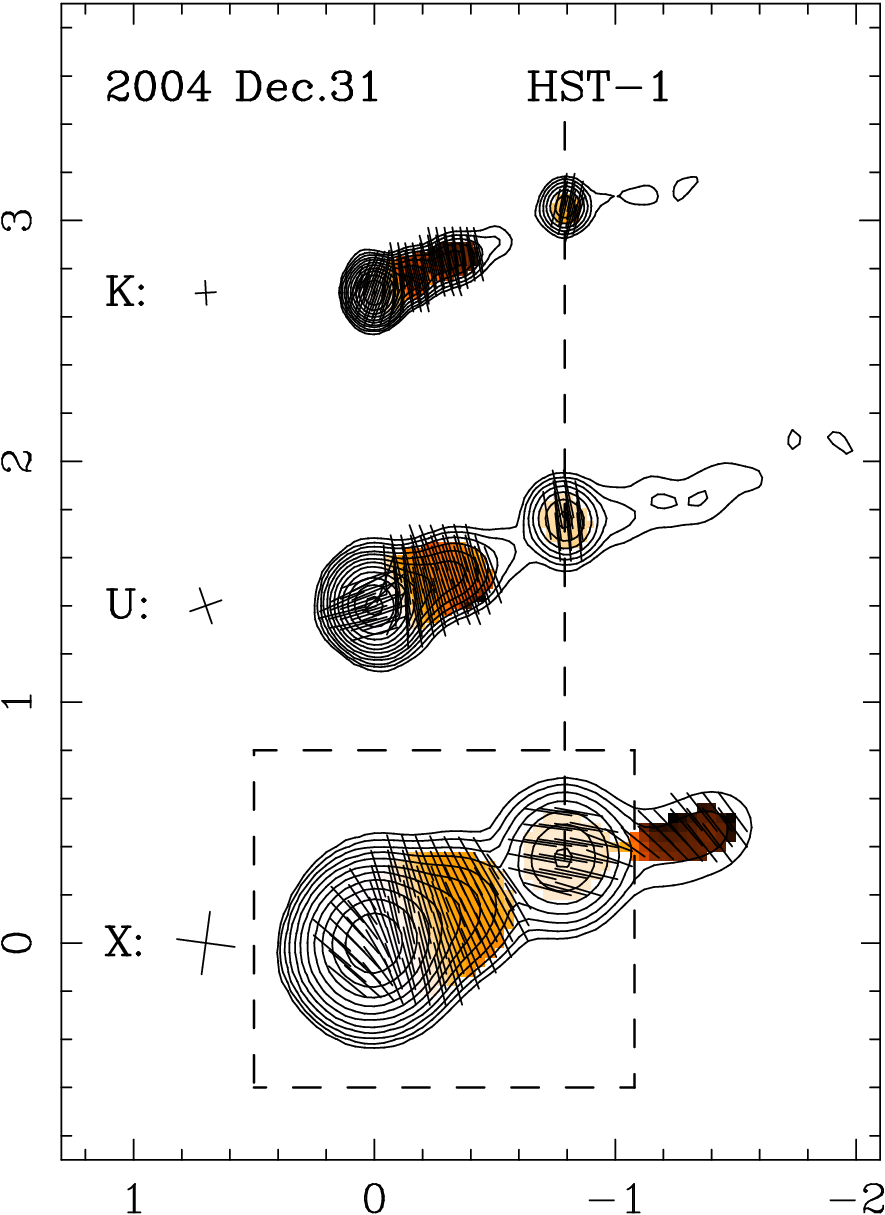}}
\caption{Total intensity contours for the central emission region of
M87 with fractional polarization (color) and EVPA vectors overlaid
at epoch 2004 Dec. 31. The contour levels start at 1.5, 1.0 \& 1.0
mJy, and increase by factors of 2, at 8 (X band), 15 (U band), and
22 GHz (K band), respectively. The corresponding restoring beams are
($0.26\times0.24$) $\rm{arcsec^2}$ at ${\rm P.A.= -7.6^\circ}$,
($0.16\times0.14$) $\rm{arcsec^2}$ at ${\rm P.A.= 19.2^\circ}$, and
($0.10\times0.09$) $\rm{arcsec^2}$ at ${\rm P.A.= 3.3^\circ}$
respectively, denoted as cross in the left of each image with its
size indicating the beam size. The locations of HST-1 lie at places
where a dashed line passes through at the 3 wavebands. The dashed-line box denotes the fitting emission region.\label{fig1}}
\end{figure}

The imaging results show that the jet extends to the northwest with a counter lobe structure clearly detected on the opposite side, and hence the detected emission region is as large as more than 50 arcseconds along the overall jet direction at 8 GHz. In this Letter, we focus only on the feature HST-1, which changes significantly in various aspects during the major outburst. Figure~\ref{fig1} shows, just as an example, the central emission region of total intensity radio structure with fractional polarization (color) and EVPA
vectors overlaid at epoch 2004 Dec. 31 at 8, 15, and 22 GHz. One can
find that the total intensity radio structures appear similar at these frequencies with a little more extended structure detected at low frequencies. The fractional polarization and EVPA changes with distances to the core of M87.

HST-1 is defined here to be the point-like emission feature excluding the extended structure downstream from HST-1, as is shown in Fig.~\ref{fig1}. When the emission region upstream of HST-1 is not taken into account, the radio structure looks like a core-jet structure extending to northwest. The EVPA of HST-1 changes apparently with frequencies, which is due to Faraday rotation from HST-1 itself and its ambient medium \citep{che10,hom09}. All quantitative analysis with epoch and frequency, including the total flux density, polarization and RM of HST-1, are given in the following.

\subsection{Flux and Spectral Variability of HST-1}

\begin{figure}
\begin{center}
\includegraphics[width=5.5cm,angle=-90]{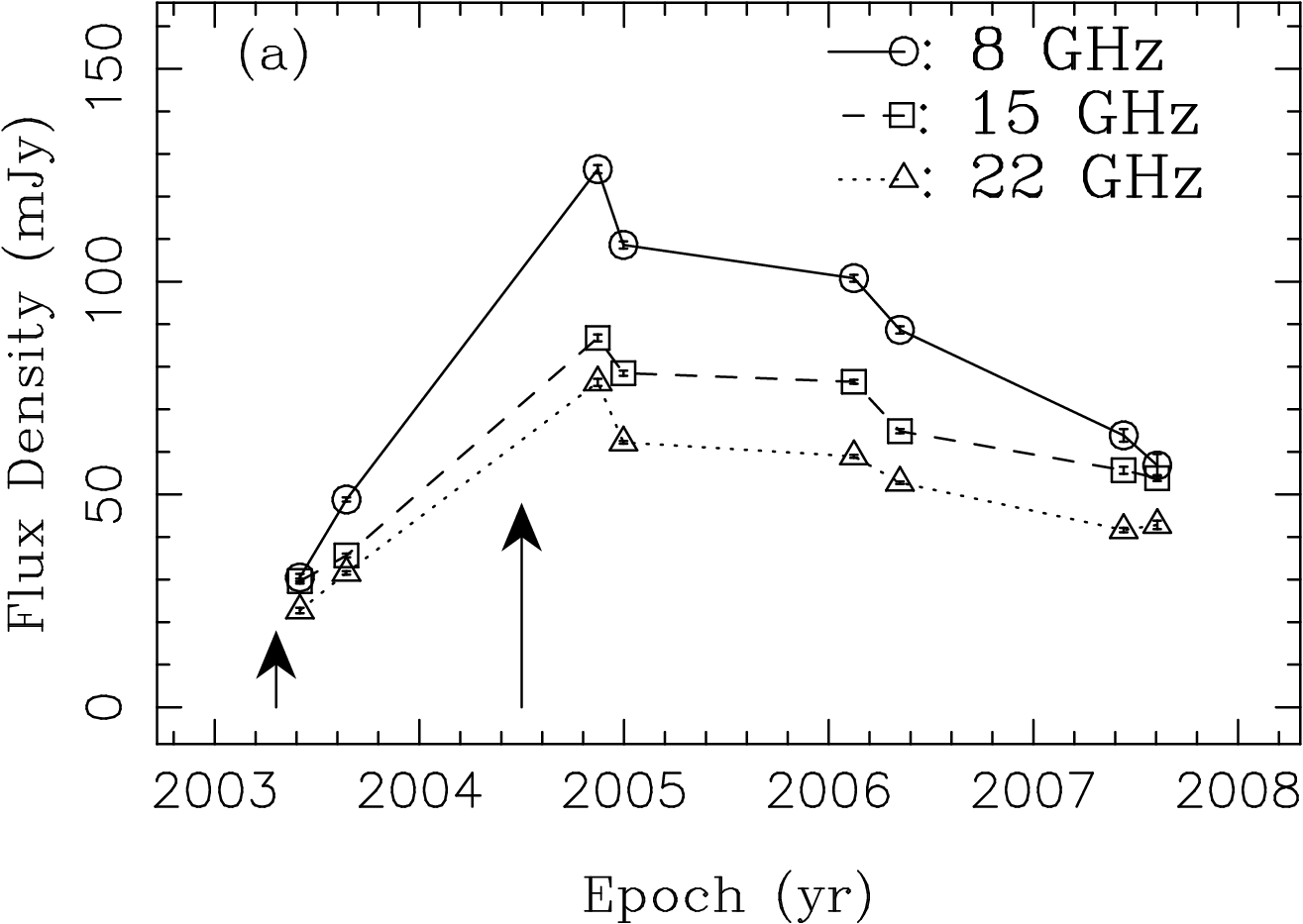}
\includegraphics[width=5.5cm,angle=-90]{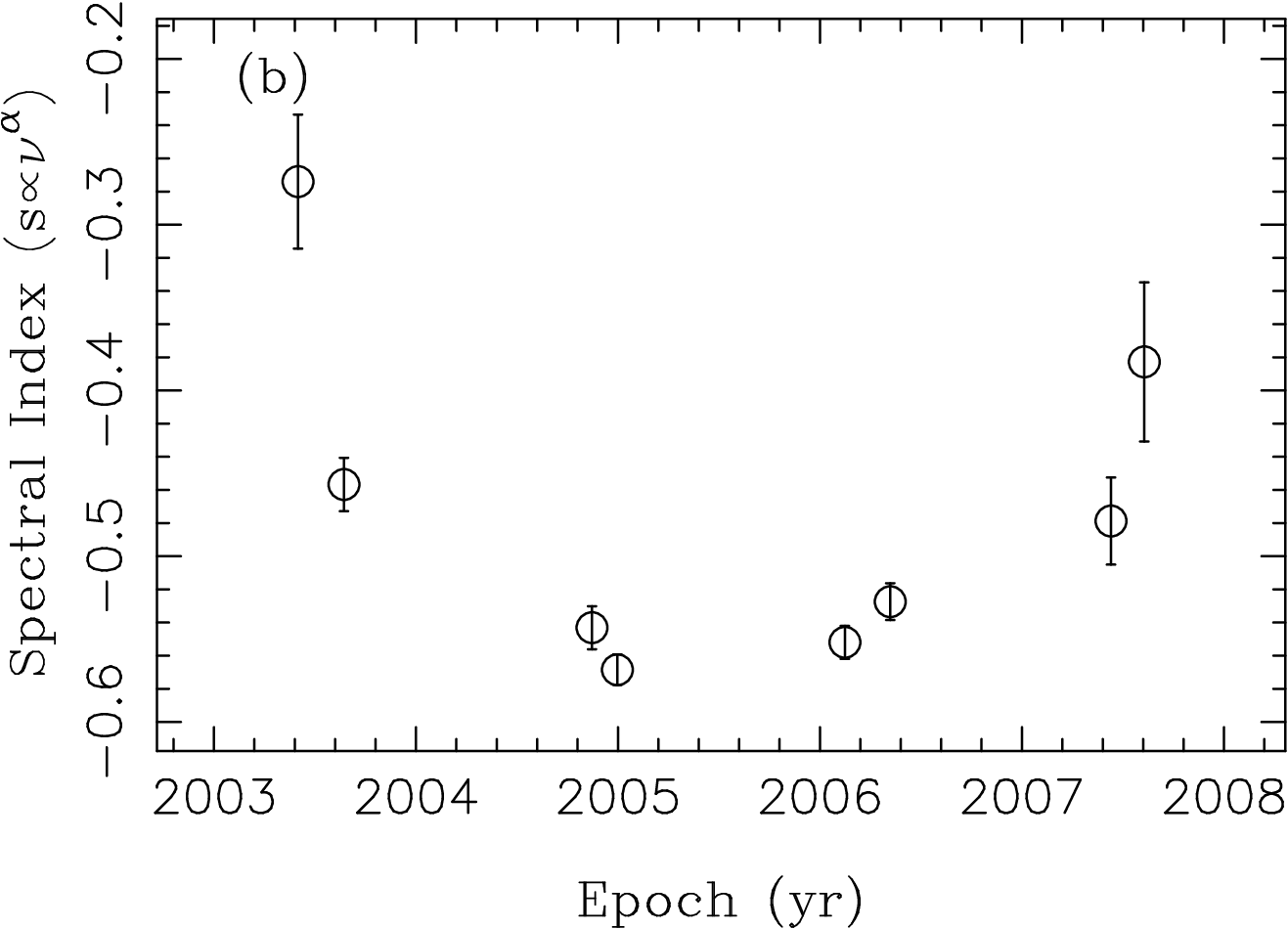}
\caption{(a) Radio light curves of HST-1 at 8 (solid line), 15
(dashed line) and 22 GHz (dotted line), the arrows indicate the
times at which two subcomponents are emerging from the stationary
one within HST-1 on mas scale derived by Cheung et al. (2007); (b) Spectral
variability of HST-1 with indices obtained through fits to the 3
waveband data. The errors to total flux density are obtained in the
course of model fitting, and the errors to spectral indices are
derived in the course of fits. \label{fig2}}
\end{center}
\end{figure}

The total flux density variations at the three wavebands are shown in Fig.~\ref{fig2}a, of which the 15 GHz flux densities are in agreement with the results for the first several epochs available in \cite{har09}. It is shown 
that the outburst grows relatively early at low frequency in the rising stage, which can also be found in a much wider frequency range from radio to X-ray \citep{har09}. This is contrary to the general case for blazars, where the flares at lower frequencies are generally delayed owning to synchrotron self-absorption (e.g., \citealt{mar08}).

Since the flux density is relatively high at low frequency at all epochs (Fig.~\ref{fig1}a), the data is fitted over the 3 wavebands in a power law spectrum at each epoch. One can find that the spectrum softens with epoch in the phase of rising flux, reaches the softest state in the peak stage, and then gets hardening in the decaying stage, as is shown in Fig.~\ref{fig2}b. During the rising stage, at least two subcomponents are emerging from the stationary subcomponent in the eastern end of HST-1 at times shown by arrows in
Fig.~\ref{fig2}a \citep{che07}, which implies that the newly emerging subcomponent(s) has a relatively steep spectrum, and hence changes the hardness of the overall spectrum of HST-1 during the outburst assuming that the spectrum keeps constant for the stationary subcomponent.

Note that the bright components of the radio jet appear to accelerate with distance from core to HST-1 with subluminal speeds, whereas several knots placed further out from the core (0.8-6.3 arcsec) move superluminally down the jet \citep{sta06}. In particular, multi-epoch optical and Very Long Baseline Array (VLBA) observations all showed that the unresolved stationary feature at the upstream edge of the HST-1 knot seems to emit various subcomponents down the jet with relativistic speed \citep{bir99, che07}. That is to say, except for the stationary subcomponent which is special due to either standing shock or optical opacity there, all other subcomponents may well be accelerated, and electrons are energized in this region. On the other hand, a major flare of HST-1 occurred during the period of 2003 to 2007, accompanied by two
subcomponents emerging from the stationary one in the rising stage
\citep{che07,har09}. Adiabatic compression at least should not be the dominant mechanism for energizing the radiating electrons, or else the spectra would be hardening in the rising stage \citep{har03}, which is contrary to the observed results. It is more likely that, as suggested by \cite{har03}, the increased intensity arises from the injection of new radiating particles that are
energized in the region of HST-1 due to the fact that no significant emission from radio to X-ray are detected immediately upstream of HST-1 even in the outburst period. Hence, there must be a way or mechanism, for example, a helical magnetic field, that plays a role in feature acceleration and particle energizing in the major outburst, which will be further discussed bellow.

\subsection{Variations of Intrinsic EVPA and Fractional Polarization}

\begin{figure}
\begin{center}
\includegraphics[height=8.0cm, angle=-90]{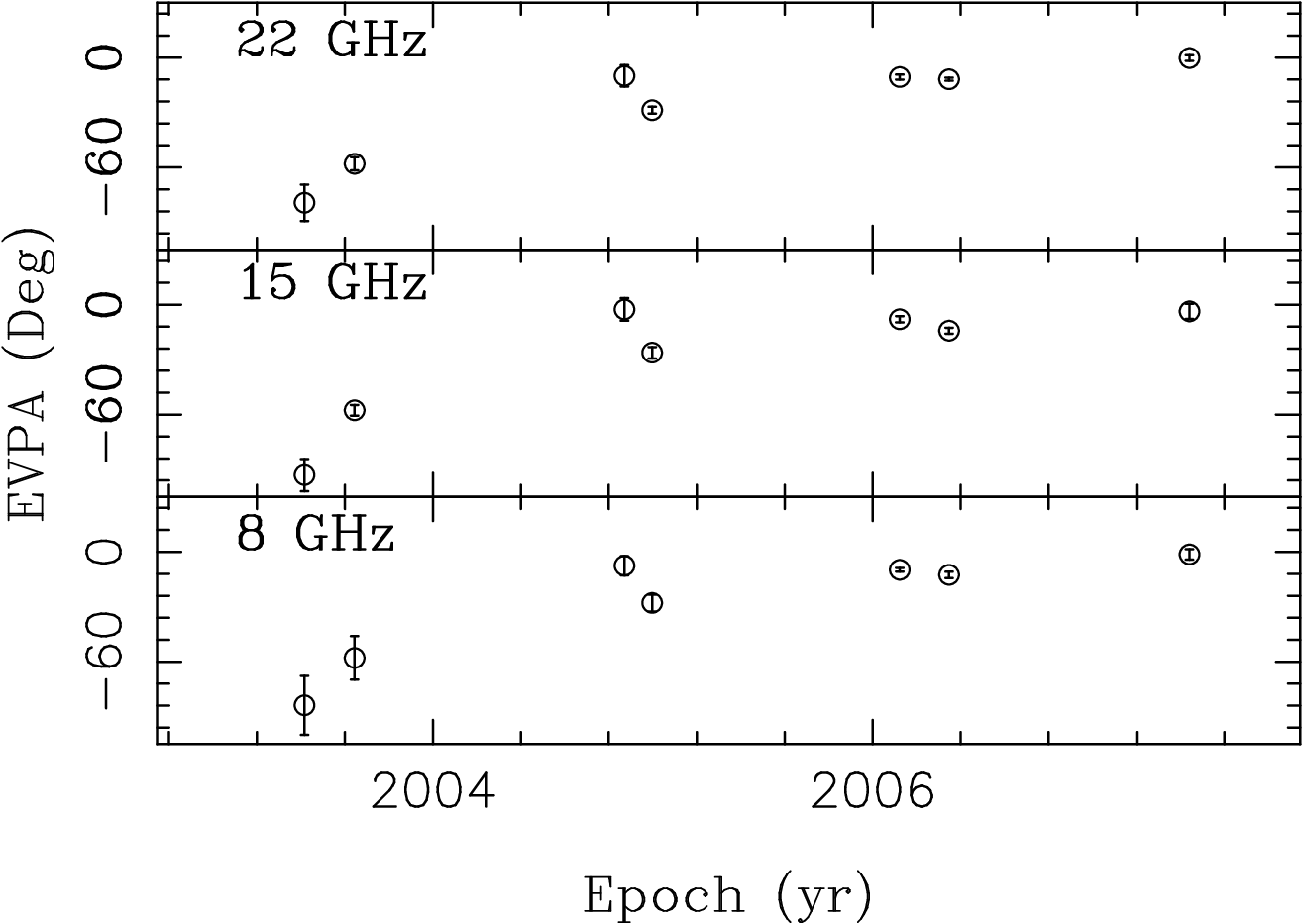}
\hspace{0.5cm}
\includegraphics[height=8.0cm, angle=-90]{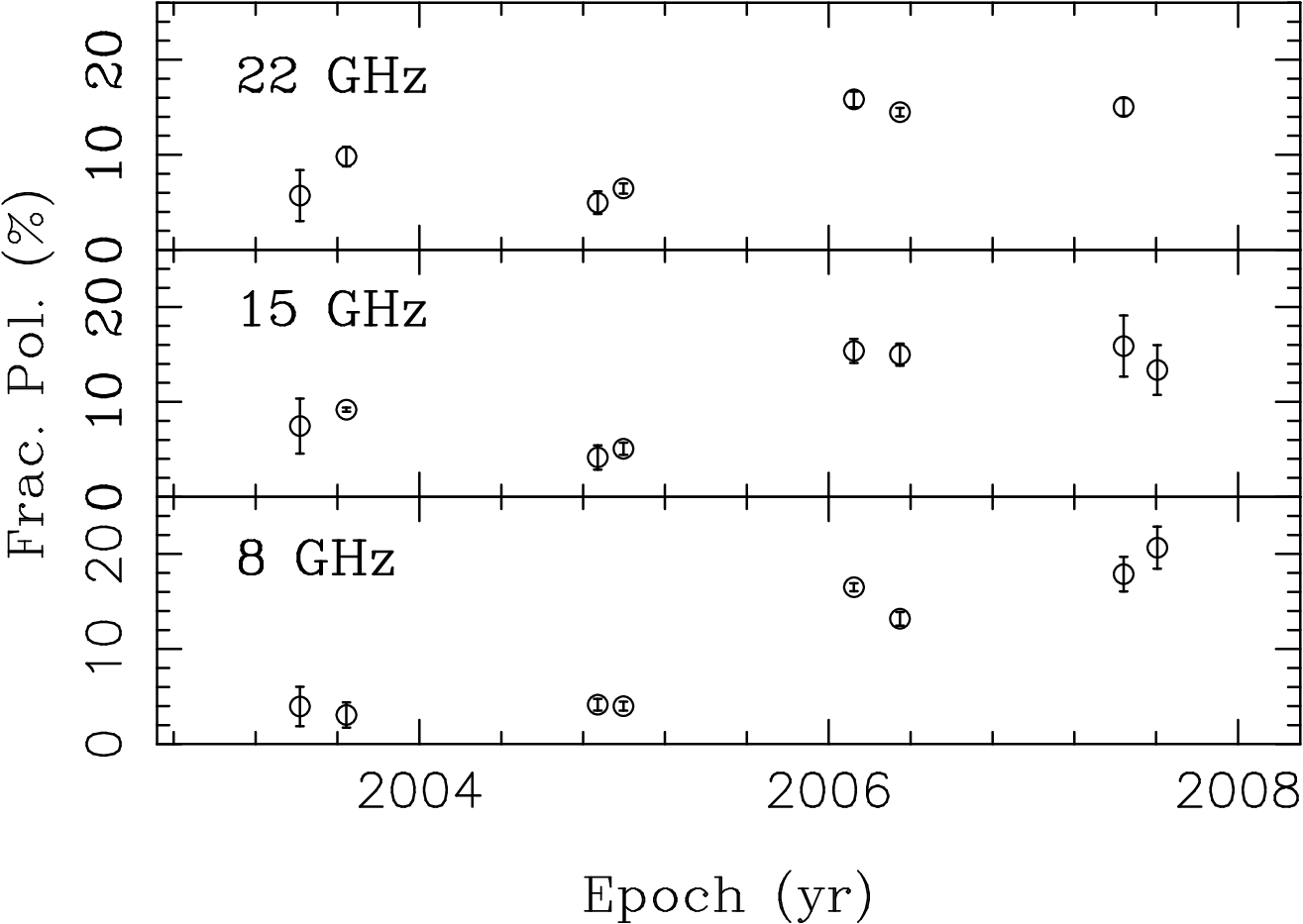}
\caption{Bottom panel: Fractional polarization of HST-1 with time at
8 (bottom), 15 (midle) and 22 GHz (top); Top panel: RM corrected
EVPA of HST-1 with epochs at 8 (bottom), 15 (middle) and 22 GHz
(top). The magnitude of error bars are 3 times of uncertainty obtained through transfer from stokes  `I', `Q', `U', and their errors.}
\label{fig3}
\end{center}
\end{figure}

According to theoretical models (e.g., \citealt{mei01}), jets in AGN are propelled and collimated by magnetic fields twisted by differential
rotation of the black hole¡¯s accretion disk or inertial-frame-
dragging ergosphere, and the primary observational indicator of magnetic collimation requiring a helical magnetic field in the spine of the jet is the evolution of the polarization, which manifests as EVPA rotation as a feature accelerates and focuses along its spiral path \citep{mar08}. The EVPA rotation has been detected in the monitoring of EVPA to the sources BL Lac and PKS 1510-089 \citep{mar08,mar10}. Through multi-epoch polarized VLA observations at 3 wavebands over 3 years, we find a similar case with the feature HST-1 in M87 jet. The intrinsic EVPA changes with time in a regular way, as is shown in the top panel of Fig.~\ref{fig3}. In this map, the Faraday rotation effect,
which will give rise to EVPA change by $\rm{\triangle\chi =
RM\lambda^2}$, has been eliminated at all the epochs. Here, RM is obtained through fits of EVPA to the wavelength square $\lambda^2$. Except for the epoch 2007 Aug. 10, at which the observational EVPAs of calibrator 3C 286, and hence M87 at 22 GHz are not available, the resultant RM-corrected EVPA shows a roughly monotonous variation from $\sim-90^\circ$ to $\sim0^\circ$. Although the monotonous variation of EVPA may be also interpreted as a random walk of the resultant polarization vector direction as simulation cells with random magnetic field orientations enter and then exit the emission region, the probability is supposed to be very low \citep{mar10}. It is more likely that the observed long-term EVPA variation is dictated by the magnetic geometry, as suggested by \cite{mar08}.

Another point is that as mentioned above, the emission features may well be accelerated in HST-1 region, which is also supported by the fact that almost all subcomponents within HST-1 move superluminally \citep{che07}, and acceleration obviously occurs for HST-1 itself over the whole major outburst period \citep{gio11}, whereas all components upstream of HST-1 are subluminal \citep{sta06}. Based on the model proposed by \cite{mar08}, the emission features will move following a spiral path if a helical magnetic field exists, and different features may follow different subset of streamlines through the acceleration and collimation zone. This scenario has been definitely detected by the multi-epoch VLBA observations of M87 during the flaring period of HST-1 by \cite{che07}. The VLBA observations show that the resolved subcomponent $a$ moves downstream at $\sim2.49c$ at $\rm{P.A.\simeq295^\circ}$ initially, and then deflect to the direction of $\sim289^\circ$ with apparent speed of $\sim1.41c$, differing to the overall jet direction of $290^\circ$. In addition, the other two subcomponents $b$ and $c$ move down the jet at $\rm{P.A.\simeq279^\circ}$, and the routes also appear not straight. It can be interpreted that different emission features may gain energy across different part of its cross-sectional area, and trace down the different subset of streamlines, and the changes in apparent speed may mainly attribute to the orientation change down the jet. \cite{gio11} presented an e-EVN image at epoch 2010 Jan. 27, which shows a significant orientation difference of the substructures within HST-1 to that at epochs by 2007 given in \cite{che07}. If the helical magnetic field around HST-1 do exists, each subcomponent may follow different subset of spiral streamlines down the jet, and with time, result in the significant orientation change of the substructures within it. All these effectively strengthens the argument that a helical magnetic field around HST-1 plays an important role in these phenomena. The approximate 100 pc scale of the disk of ionized gas observed in M87 by HST \citep{for94} may indicate to some extent the presence of helical magnetic fields around HST-1, which is also about 100 pc away from the active centre \citep{sta06}.

Fractional polarization reflects the degree of magnetic field order, which may be affected by Faraday rotation and the medium opacity in the passage to the observer \citep{hom09}, and polarization level variability is argued to arise from the source itself as well as its immediate surrounding medium \citep{hom09}. We show the variations of the fractional polarization in the bottom panel of Fig.~\ref{fig3}, which shows that the fractional polarization varies in a quite complex way. However, it is quite affirmative that the fractional polarization roughly minimizes in the peaking stage, and is relatively low in the rising stage, and high in the decaying stage. That is to say that the depolarization effect due to Faraday rotation and medium opacity is relatively strong in the rising stage of the major outburst.

\subsection{Faraday Rotation of Emission from HST-1}

\begin{figure}
\centerline{\includegraphics[width=6cm,angle=-90]{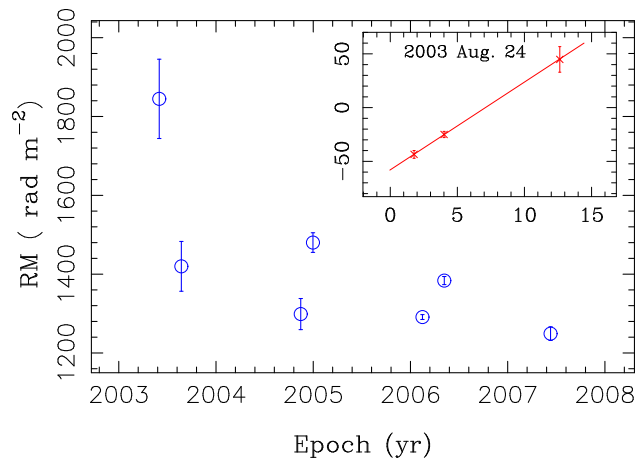}}
\caption{Rotation measure of HST-1 with time. The top right inset is
a ${\rm\lambda^2}$ regression to the observed Faraday rotation at
epoch 2003 Aug. 24 with abscissa ${\rm\lambda^2}$ in ${\rm cm^2}$
and ordinate EVPA in deg. The magnitude of error bars in the inset
are 3 times of uncertainty.} \label{fig4}
\end{figure}

Faraday rotation is often detected from jet emission in radio-loud AGNs. \cite{zav04,zav05} suggested that it occurs external to the jet by excluding several other identities such as broad and narrow line regions, and giving evidence of high fractional polarization to rule out internal Faraday rotation in 3C 273, while internal Faraday rotation gets supported in model-fitting to the full polarization spectra of 3 jet components simultaneously in a consistent physical picture in 3C 279 \citep{hom09}. The rotation measure can be expressed as $\rm{RM\propto\int N(\textbf{s})\textbf{B}\cdot d\textbf{s}}$, where N is the electron number density, \textbf{B} is the magnetic field, and \textbf{s} is the emission path to the observer. Obviously, RM may provide constraints on number density and magnetic field in the passage, which is here obtained by fitting the observed EVPAs over 3 wavebands with Faraday wavelength square law. The RM variation with time for HST-1 is shown in Fig.~\ref{fig4}, in which the top right inset gives the fit of the observed EVPAs in deg to the squared wavelength in $\rm{cm^2}$, as an example, at epoch 2003 Aug. 24. One
can find that the EVPAs obey the $\lambda^2$ Faraday rotation law well at the epoch. This means that although HST-1 contains complex substructures, its EVPA still obeys the Faraday rotation law well.

Figure~\ref{fig4} shows that RM has the largest value in the beginning, and gradually decreases with fluctuation present due to subflares occurred in the mean time \citep{har09}. This is consistent with the fact that fractional polarization is relatively low in the early stage of the outburst, and as expected, the RM variability may well arise from the interior of HST-1. Faraday rotation may give rise to depolarization by a factor
of $\rm{sin(\Phi)/\Phi}$ \citep{bur66, hom09}, where
$\rm{\Phi=2\Delta\chi}$, if Faraday rotation occurs internal to HST-1. According to the resultant RM values shown in Fig.~\ref{fig4}, the dispersion effect is supposed to be quite significant at 8 GHz in comparison to that at the other two frequencies if Faraday rotation occurs internal to HST-1, for
example, the polarization level at 8 GHz would be 4 times lower than
that at 15 GHz even when RM took a relatively small value of $\rm{1000~rad~m^{-2}}$. Moreover, the relatively large beam size at low frequencies would make the fractional polarization even lower if it had an effect on the polarization level \citep{tay98}. However, one can find in the bottom panels of Fig.~\ref{fig3} that after 2006, the polarization levels are almost comparable at the 3 frequencies. This implies that, although the RM variation is attributed to the plasma internal to HST-1, quite a large part of Faraday rotation may occur external to HST-1. Especially in the decaying stage, the Faraday rotation may occur dominantly external to HST-1, as is required by the comparable fractional polarization at the 3 wavebands.  This is consistent with the fact that RM decrease gets slower with time, and more importantly, is supported by the presence of diffuse emission around HST-1 measured with VLA at low frequencies, as is shown in Fig.~\ref{fig1} of \cite{har08}.

\section{Summary}

We reduced 8 epoch polarimetric observations of M87 with VLA A array at 8, 15, and 22 GHz from 2003 to 2007. Due to the significant blazar-like nature shown in HST-1 \citep{har08}, the flux density, spectral, polarization, and RM variabilities are investigated in the period of a major outburst, during which a $\gamma$-ray flare is detected to be associated with it.

Contrary to the general case for blazars, the outburst of HST-1 grows relatively early at low frequency, which is consistent with the spectral variability that the spectrum is softening in the rising stage. This implies that the newly emerging subcomponents have relatively steep spectra, and hence make the overall spectrum softening. In addition, these newly emerging subcomponents from 2003 to 2004 are superluminal \citep{che07}, and hence expected to be accelerated in this region due to the fact that all components
upstream of HST-1 are subluminal, and become faster with distance from the core.

The RM-corrected EVPA of HST-1 shows a monotonous variation from
$\sim-90^\circ$ to $\sim0^\circ$ at the 3 wavebands over a period of more than 3 years, and in this period the subcomponents within HST-1 move down along the routes deviating from the overall jet direction with the motion of one subcomponent bending from one side of the large scale jet axis to another, and another two subcomponents deflecting significantly from the jet axis. These give quite strong evidence for the presence of a helical magnetic field around HST-1, which also gets support from the fact that the emission features may get accelerated in this region. The fractional polarization varies with epochs, and is relatively low in the rising stage in comparison to that in the decaying stage. This implies that the internal Faraday rotation and medium opacity play an important role to the variation of polarization level during the outburst. Moreover, in the decaying stage after 2006, the polarization levels are almost comparable at the 3 wavebands, indicating that Faraday rotation occurs dominantly external to HST-1 in this phase, or else it would be much lower at 8 GHz than measured.

\section*{Acknowledgments}

This work was supported in part by the grants 10625314, 10633010 and
10821302, KJCX2-YW-T03, 2007CB815405, the CAS/SAFEA International
Partnership Program for Creative Research Teams, and the Science and
Technology Commission of Shanghai Municipality (09ZR1437400).

\label{lastpage}

\end{document}